\documentclass[final,5p,times]{elsarticle}
\usepackage{amssymb}
\usepackage{xcolor}
\usepackage{xparse}
\usepackage{hyperref}
\usepackage{xurl}

\hypersetup{
    colorlinks,
    linkcolor={red!50!black},
    citecolor={blue!50!black},
    urlcolor={blue!80!black}
}

\NewDocumentCommand{\todo}{o}{%
    \textcolor{red}{TODO%
    \IfValueT{#1}{: #1}}%
}

\journal{Future Generation Computer Systems}

\begin{document}

\begin{frontmatter}

\title{Onedata4Sci: Life science data management solution based on Onedata}

\author[ceitec,ncbr,uvt]{Tomáš Svoboda\corref{first}}
\author[ceitec,ncbr,uvt]{Tomáš Raček\corref{first}}
\author[ncbr]{Josef Handl}
\author[ceitec,uvt]{Jozef Sabo}
\author[ceitec,ncbr,uvt]{Adrián Rošinec}
\author[agh]{Łukasz Opioła}
\author[ceitec]{Wojciech Jesionek}
\author[ceitec]{Milan Ešner}
\author[ceitec,ncbr]{Markéta Pernisová}
\author[ceitec]{Natallia Madzia Valasevich}
\author[uvt]{Aleš Křenek\corref{corr}}
\author[ceitec,ncbr]{Radka Svobodová\corref{corr}}

\cortext[first]{Shared first authorship}
\cortext[corr]{Shared correspondence authorship}

\affiliation[ceitec]{organization={CEITEC -- Central European Institute of Technology},
            addressline={Kamenice 753/5}, 
            city={Brno},
            postcode={625 00}, 
            country={Czech Republic}}

\affiliation[ncbr]{organization={National Centre for Biomolecular Research, Faculty of Science, Masaryk University},
            addressline={Kamenice 753/5}, 
            city={Brno},
            postcode={625 00}, 
            country={Czech Republic}}

\affiliation[uvt]{organization={Institute of Computer Science, Masaryk University},
            addressline={Botanická 68a}, 
            city={Brno},
            postcode={602 00}, 
            country={Czech Republic}}

\affiliation[agh]{organization={ACC Cyfronet AGH, AGH University of Science and Technology},
            addressline={Nawojki 11}, 
            city={Kraków},
            postcode={30-950}, 
            country={Poland}}

\begin{abstract}
Life-science experimental methods generate vast and ever-increasing volumes of data, which provide highly valuable research resources. However, management of these data is nontrivial and applicable software solutions are currently subject to intensive development. The solutions mainly fall into one of the two groups: general data management systems (e.g. Onedata, iRODS, B2SHARE, CERNBox) or very specialised data management solutions (e.g. solutions for biomolecular simulation data, biological imaging data, genomic data).

To bridge this gap between them, we provide \textit{Onedata4Sci}, a prototype data management solution, which is focused on the management of life science data and covers four key steps of the data life cycle, i.e. data acquisition, user access, computational processing and archiving. \textit{Onedata4Sci} is based on the Onedata data management system. It is written in Python, fully containerised, with the support for processing the stored data in Kubernetes. The applicability of \textit{Onedata4Sci} is shown in three distinct use cases -- plant imaging data, cellular imaging data, and cryo-electron microscopy data. Despite the use cases covering very different types of data and user patterns, \textit{Onedata4Sci} demonstrated an ability to successfully handle all these conditions. Complete source codes of \textit{Onedata4Sci} are available on GitHub (\url{https://github.com/CERIT-SC/onedata4sci}), and its documentation and manual for installation are also provided.
\end{abstract}


\end{frontmatter}

\section{Introduction}
\label{sec:introduction}

Life-science experimental methods, such as cryo-EM, biological imaging methods, molecular simulations, and sequencing, generate vast and ever-increasing volumes of data. This valuable data has led to numerous breakthrough discoveries in these fields, e.g. in structural biology ( \url{https://pdb101.rcsb.org/learn/other-resources/structural-biology-and-nobel-prizes}). However, acquiring this data entails significant effort and financial resources. Consequently, storing the data for further analysis becomes imperative, ensuring reproducibility of results and facilitating data sharing within the research community. Research groups, particularly core facilities (CFs) focusing on specific experimental techniques, require an efficient data management solution to address these challenges.

Although several data management systems, such as Onedata \cite{orzechowski2023indexing}, iRODS \cite{rajasekar2010irods}, B2SHARE \cite{ardestani2015b2share}, or CERNBox \cite{mascetti2015cernbox+} are currently available, they primarily offer rather low-level functionality, and not all the features required for broader user adoption are well-matured. Adapting these generic, data-agnostic systems to specific life-science use cases poses significant challenges for researchers.

For this reason, many attempts have been made to provide a solution for a specific field. High energy physics is one of the first research areas to start emphasising the importance of handling the data deluge expected to emerge after the launch of the Large Hadron Collider at CERN in the early 2000s. Related developments produced one of the first complex distributed scientific data management services, the gLite Middleware \cite{erwin2006programming, stewart2007storage}, later integrated into the EMI middleware stack \cite{aiftimiei2012towards}. Higher-level services were built on top of these developments later. The experience in this field is summarised in a recent book \cite{barisits2019rucio}. Simultaneous developments also appeared in the US, e.g. \cite{illingworth2014data}. 

In life science domain, other systems were developed, e.g. solutions for biomolecular simulation data \cite{thibault2013ibiomes, thibault2015development}, biological imaging data \cite{lopez2019cloud}, genomic data \cite{chiang2011implementing}, multi-omics data \cite{nieminen2023sodar}, microbiome data \cite{youens2019imicrobe}, plant phenotyping data \cite{neveu2019dealing}, or environmental data \cite{aguilar2020datacloud}. But they present the other extreme -- they are highly specialised, focused only on some particular life-science discipline and non-transferable to other fields.

Motivated by these needs, we present a data management solution that caters to various use cases within the life-science domain. Still, it is not tightly connected to any particular data type as opposed to the specialised solutions mentioned above. In our work, we rather focused on common usage patterns shared across different fields. Furthermore, compared to traditional low-level data management solutions, we provide the necessary automation with reasonable configuration and user access, all available within the federated environments without any centralised authority. As the basis, we chose the Onedata data management system since it is long-term supported by EGI \cite{viljoen2016towards} and shows a potential for application in life sciences\cite{aguilar2020datacloud, yuan2020bioinformatics}. Here, we introduce its extension \emph{Onedata4Sci}.

\section{Architecture of Onedata}
\label{sec:architecture_onedata}

\begin{figure}[ht!]
\includegraphics[width=\linewidth]{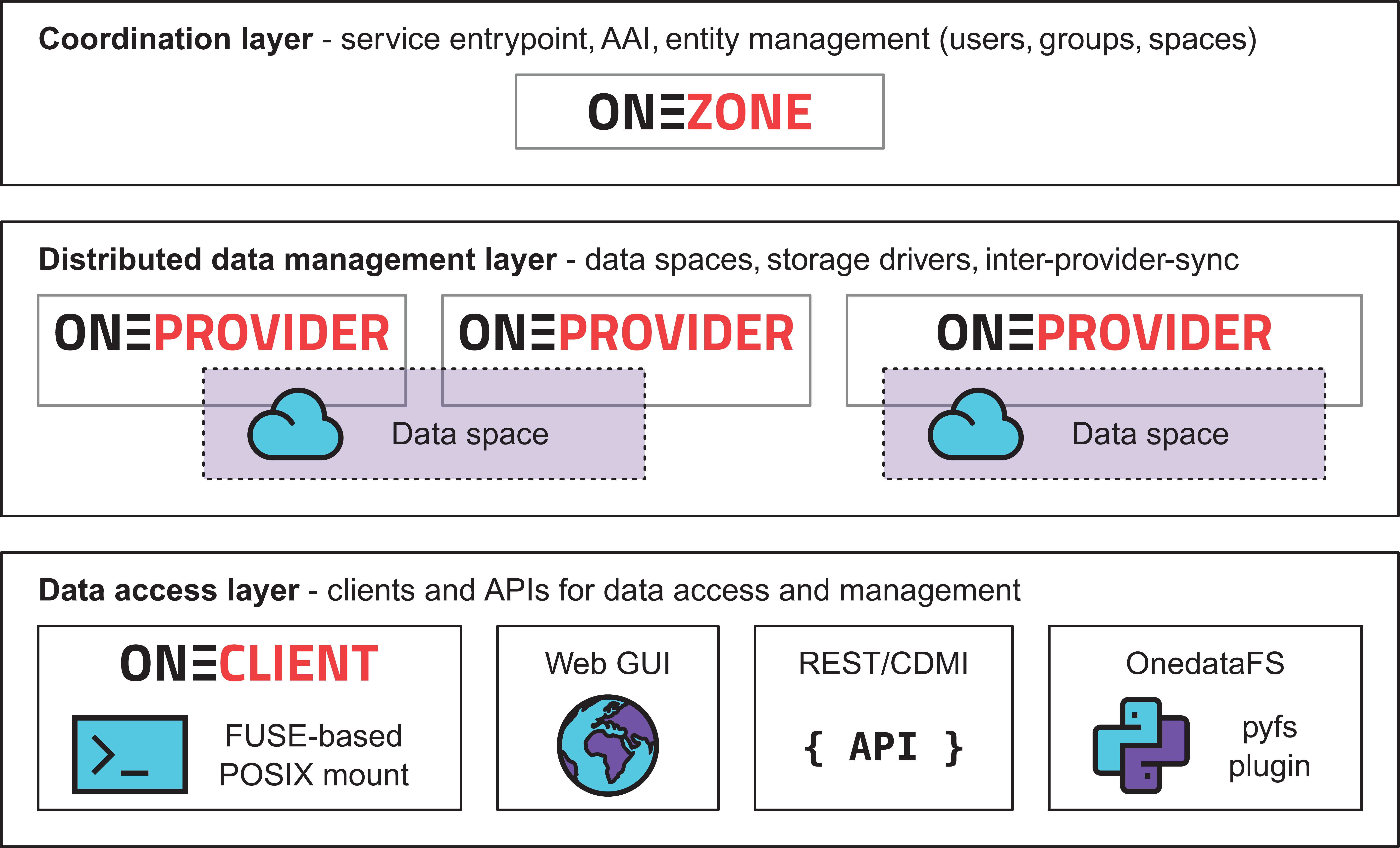}
\caption{Onedata architecture consists of Onezone, Oneprovider, and different clients for data access and management.}
\label{fig:onedata_overview}
\end{figure}

Onedata (see an architecture overview in Figure~\ref{fig:onedata_overview}) is a high-performance data management solution that offers unified data access across globally distributed environments, exposing a global virtual POSIX-like filesystem. It is built around the concept of \emph{data spaces} -- logical data containers that integrate access to physical data located on different (types) of storage systems in different data providers. In essence, its architecture consists of three types of components: Onezone, Oneprovider, and clients.

\begin{description}
    \item[Onezone] is a service that manages a Onedata \emph{zone}, i.e. an environment comprising an arbitrary number of independent data providers. However, those must accept Onezone as a trusted authority that mediates their communication and manages high-level entities in the system (e.g. users, groups, data spaces). Onezone serves as an entry point to the system, providing a common overlay platform over all the storage resources. End users are authenticated with their federated identities~\cite{oidc}.
    \item[Oneprovider] is a service deployed on site of a data provider that implements distributed data management in logical data spaces by virtualising access to different physical storage back ends (e.g. POSIX, Ceph, S3). It communicates with other Oneproviders as needed under Onezone's mediation in order to synchronise the metadata of distributed data sets or perform data transfers. Oneprovider acts as a server for different types of clients.
    \item[Various clients,] besides a standard web GUI, can be used to manage and access the data; REST API and PyFilesystem plugin OnedataFS are provided. Moreover, Oneclient can mount the Onedata filesystem locally, either on personal computers or worker nodes in computing centres, and leverage \emph{DirectIO} for optimised access to the storage back end (e.g.\ S3).
\end{description}

\section{Architecture of Onedata4Sci}
\label{sec:architecture_onedata4sci}

\begin{figure}[ht!]
\includegraphics[width=\linewidth]{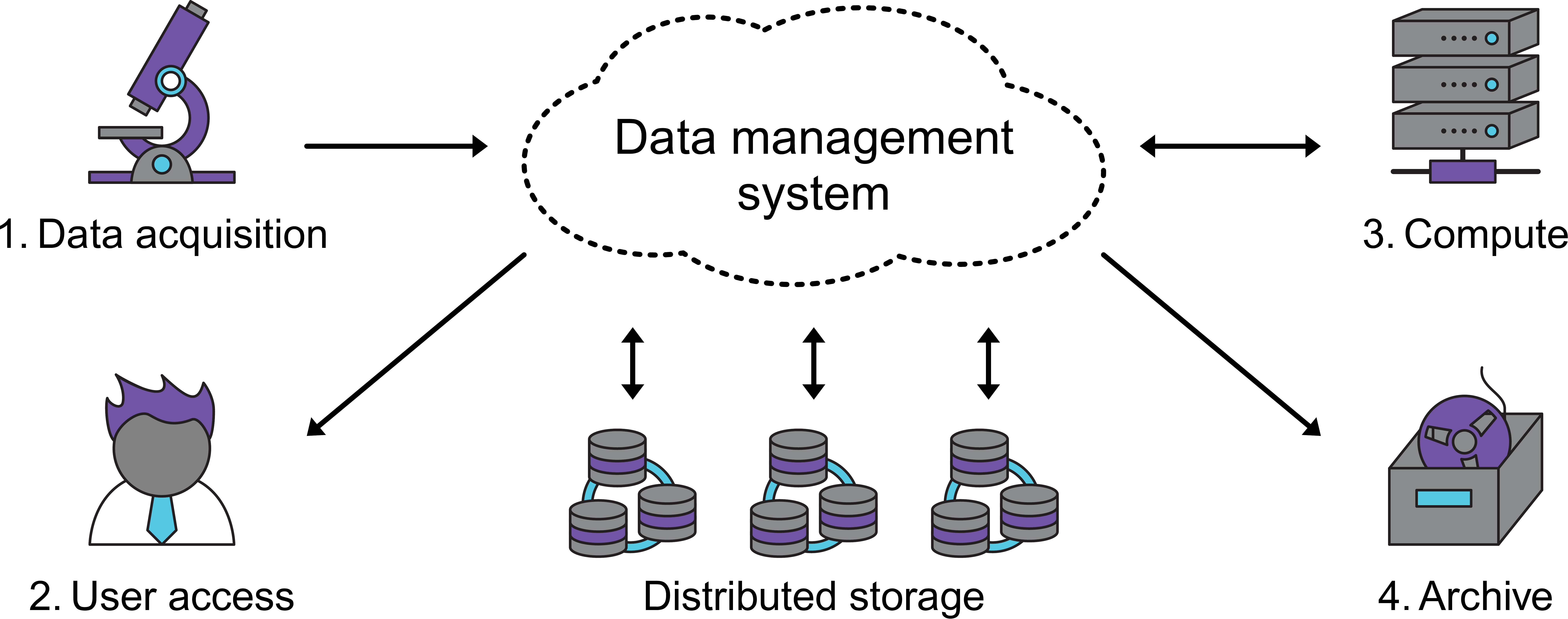}
\caption{Common workflow pattern of data management solutions}
\label{fig:data_workflow}
\end{figure}

At the top-level view, virtually all feasible use cases (in life sciences and also in other domains) share a common workflow pattern (see Figure~\ref{fig:data_workflow}). Experimental data are acquired and stored (in more or less raw form), the datasets are assigned their associated metadata and permanent identifiers, and they enter the data management system. Further, the users access and share the data. In some scenarios, the data are also processed computationally so that secondary datasets are produced and shared in a similar way. Finally, all the data can be archived on the long-term storage. 

The general data workflow described in the previous paragraph serves as a basis for the architecture of \textit{Onedata4Sci}. Each of the four steps of the data life cycle will described in detail, from the initial acquisition to the long-term archival.

\subsection{Data acquisition}
\label{ssec:data_acquisition}

Onedata offers so-called storage import, which essentially monitors a storage system location with the direct scan mechanism, importing the data present there into a Onedata space (see Section~\ref{sec:implementation} below for details). Depending on various conditions and requirements, two scenarios are possible:

\subsubsection{Mounted storage}
\label{sssec:mounted_storage}

\begin{figure}[ht!]
\includegraphics[width=\linewidth]{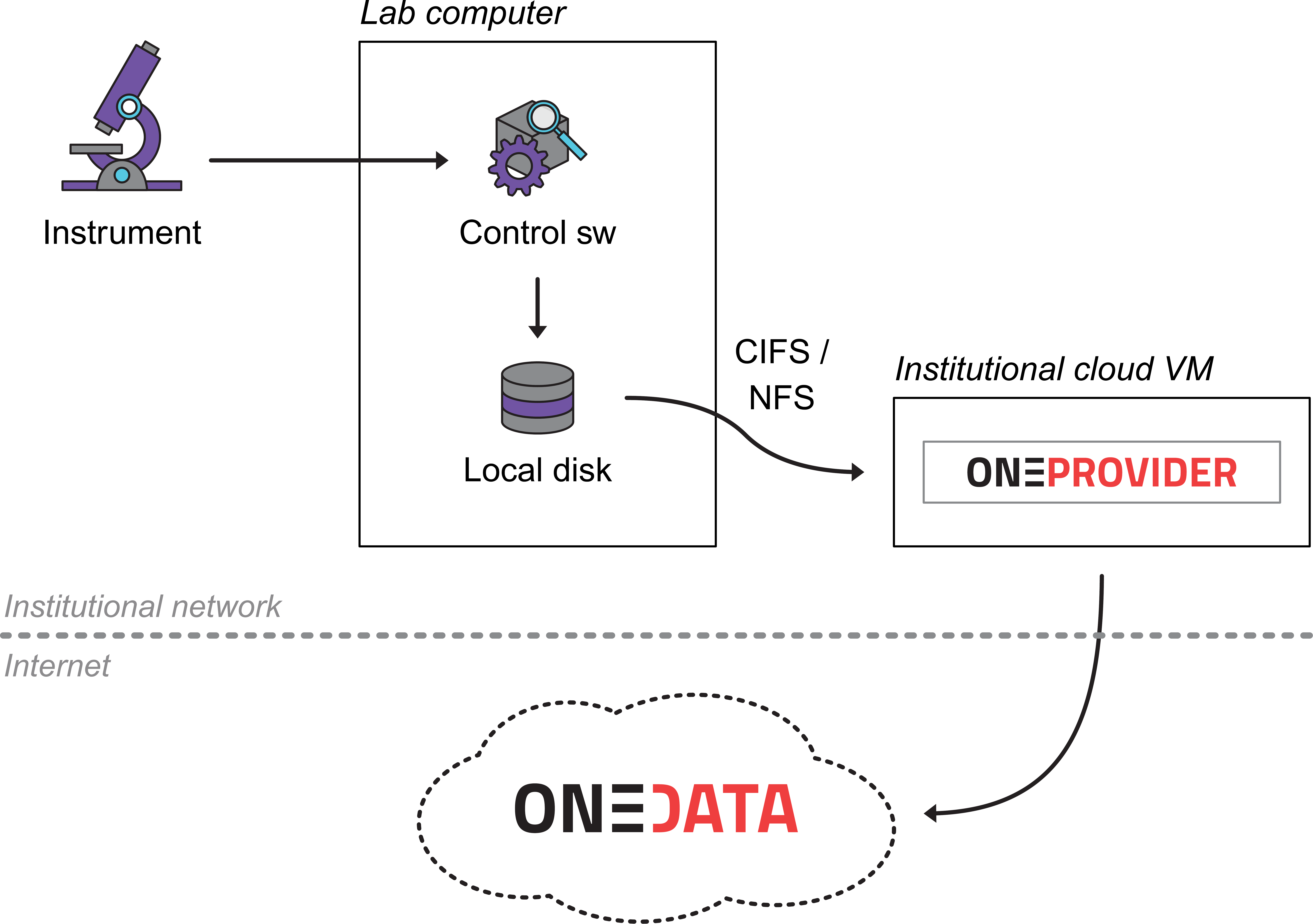}
\caption{Mounted storage pattern}
\label{fig:mounted_storage}
\end{figure}

Data storage is tightly coupled with the data acquisition process. Typically, the instrument is connected to a computer hosting the data storage (Figure~\ref{fig:mounted_storage}). The computer runs proprietary software to control the instrument; the operating system, usually Windows, could have been customised, or an outdated version could have been used. Further software installation is undesirable (or even forbidden by the instrument vendor). Typically, during the acquisition, the data must be stored on a local disk, not a remote network drive, in order to prevent data loss in case of a temporary network outage.

In this scenario, we leave the instrument control computer intact, and the only requirement is exposing the volume where the data are stored as a network drive (using CIFS or NFS protocols). The Oneprovider component is installed elsewhere, on a dedicated computer or a virtual machine, the data volume is mounted there, and Oneprovider is configured to monitor this filesystem.

Minimal changes are required for the original standalone setup. On the other hand, this configuration is limited to moderate data sizes and throughput only (at most hundreds of gigabytes per day). The bottleneck is the filesystem export by a computer which is not optimised for this task.

\subsubsection{Oneprovider on-site}
\label{sssec:oneprovider_onsite}

\begin{figure}[ht!]
\includegraphics[width=\linewidth]{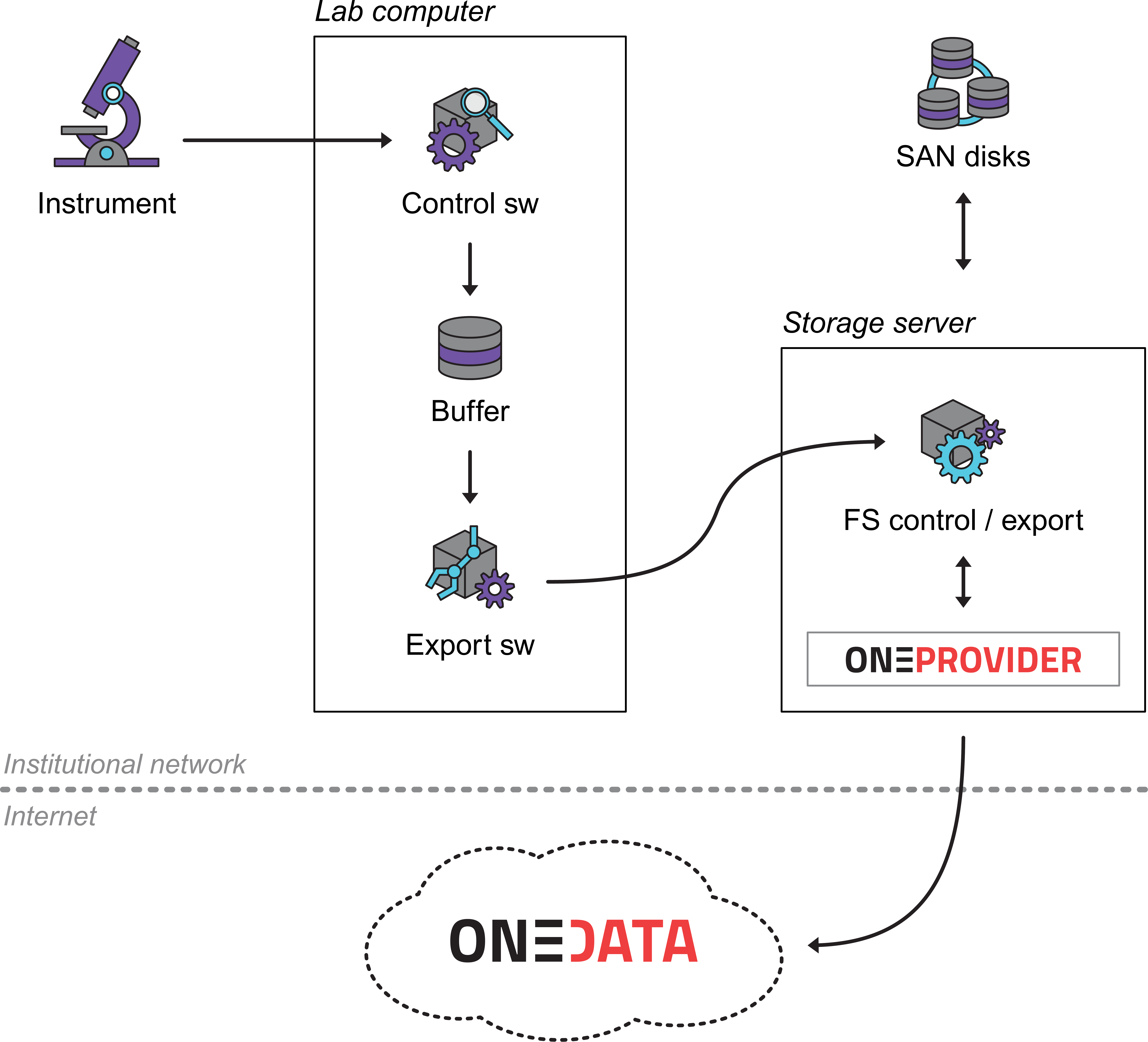}
\caption{Provider on-site pattern}
\label{fig:provider_onsite}
\end{figure}

Data storage is set up as a dedicated system, e.g. SAN devices attached to a cluster of computers implementing a distributed file system (Figure~\ref{fig:provider_onsite}). The data volume is re-exported (with NFS of CIFS again) in read-write mode and is mounted to the instrument control computer. Local storage is used for buffering the acquired experimental data only, and modified or extended acquisition software is responsible for copying the resulting datasets to the remote data volume.

The Oneprovider component is set up on the storage server. Therefore, it monitors its local (or SAN, which makes little difference) filesystem.
This scenario allows significantly higher data volumes and throughput (we tested it with terabytes per day) because of the dedicated, highly optimised storage system. However, more flexibility (extensions or modifications of the software) is required on the data acquisition computer.

\subsection{User access}
\label{ssec:user_access}

Four different cases are recognised to access the data stored within the platform.

\subsubsection{Web interface}
\label{sssec:web_interface}

After the data are acquired (or even while acquisition is in progress), the user receives a web link where the dataset can be browsed in a basic way and individual files retrieved, as provided by Onedata. The link is desired to be permanent, i.e. the particular dataset can be accessed via the link even after a long time after it has been moved to archival storage.

\subsubsection{Access control with user groups}
\label{sssec:user_groups}

The datasets can be accessed by sharing the initial link. However, this is not the recommended method – the data owner loses full control of access to the data, and whoever has the link can read them. Therefore, we create a specific access group in Onedata for each dataset; the data owner can control the group memberships, and individual users are authenticated with their federated identities. Moreover, other groups with less ad-hoc controlled membership (e.g. ``all members of my lab'' managed by the HR department) can be propagated to Onedata and granted access to the datasets too.

\subsubsection{Bulk download}
\label{sssec:bulk_download}

The basic web interface may not be suitable for managing datasets with larger numbers of files. Therefore, we also provide an option for bulk downloading a zip archive of all files. At the time of writing this manuscript, this functionality is available as a Python script (\url{https://github.com/CERIT-SC/onedata-downloader}) using the Onedata API. Similar functionality is integrated into Onedata release 21.02. 

\subsubsection{Dataset mount}
\label{sssec:dataset_mount}

Finally, the user may use the Oneclient package to mount a specific dataset as a network drive on a laptop, desktop or server. Details are given elsewhere (\url{https://onedata.org/#/home/documentation/stable/doc/using_onedata/oneclient.html}).

\subsection{Computational processing}
\label{ssec:computational_processing}

In some scenarios, processing the data requires large-scale computation to be run at a computing centre. For this setup, we embrace the DirectIO operating mode of Oneclient, which mounts a specific dataset in the local filesystem at the computing node(s), while the data are accessed directly at the storage back end (S3), hence with maximal efficiency.

Depending on the data access pattern, individual files can be read directly from the mounted filesystem, or the whole dataset can be pre-staged on a local storage before starting the computation.

For the specific case of the Kubernetes computing environment, we provide a custom Kubernetes CSI driver (Section~\ref{sec:implcomp}) to optimise both performance and sharing the data among multiple computational nodes even further.

\subsection{Archival}
\label{ssec:archival}

\begin{figure}[ht!]
\includegraphics[width=\linewidth]{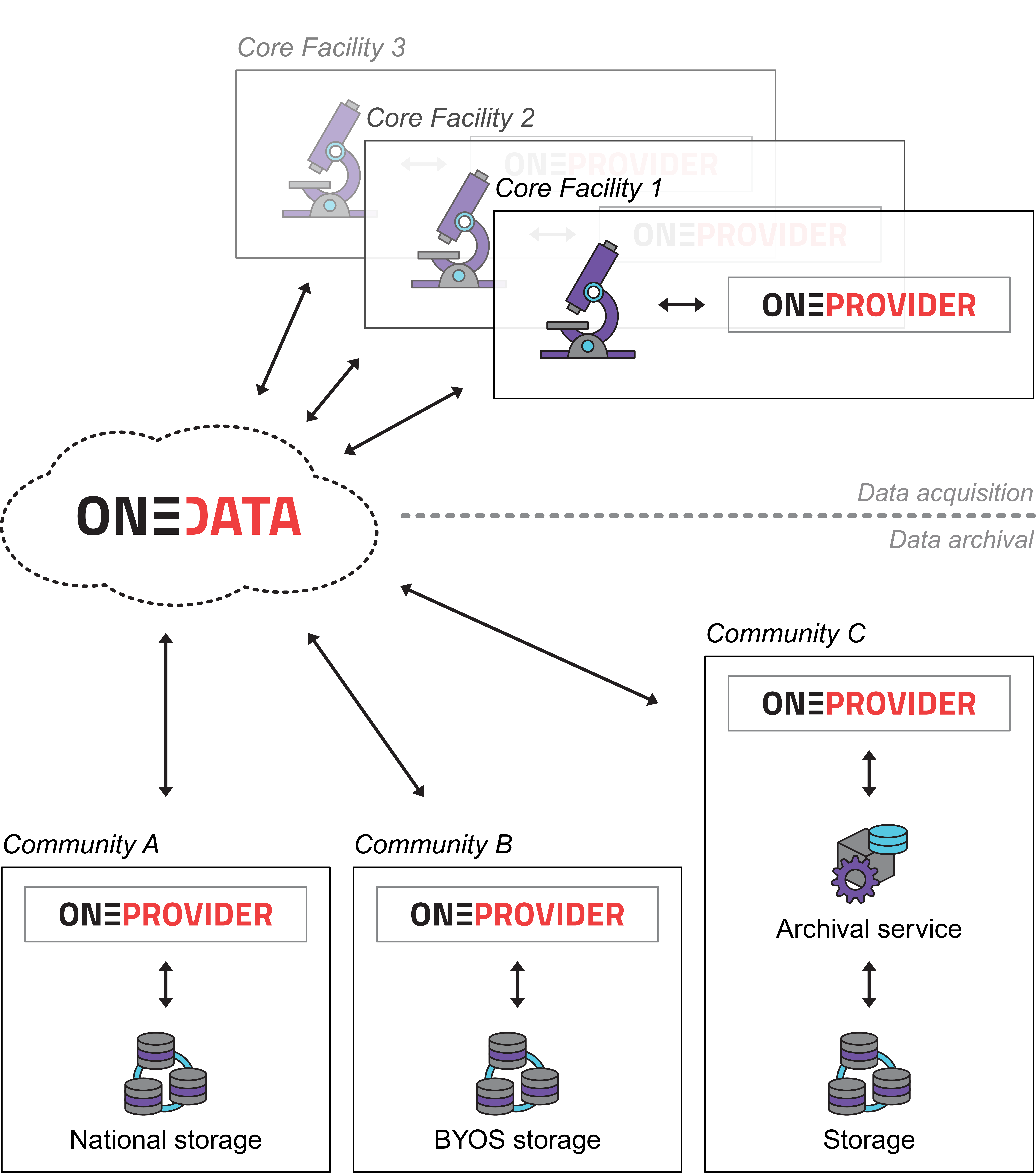}
\caption{Long-term data archival options. Acquired data (top of the figure) propagate through Onedata to the archival storage. User communities may negotiate with national data storage providers to support them (A), provide the storage service on their own (B), or leverage the integration of Onedata with an existing community archive (C)}
\label{fig:archival}
\end{figure}

With the increasing spread of open-science principles and reproducibility requirements, storing even the raw experimental data for the long term becomes a must. On the other hand, achieving these goals does not come for free, both in terms of purchasing storage resources and properly managing the data. We leave the former to the research funding agencies, while our work contributes to the latter. At the technical level, we support three slightly different scenarios (Figure~\ref{fig:archival}).

Generic storage resources provided, e.g. by national scientific e-infrastructures, can be used. The storage provider must configure a Oneprovider service with an appropriate storage back end (S3 typically) in a standard way and register it at the managing Onezone.
When the storage resources are provided by the scientific community (a.k.a. ``bring your own storage'' model), the technical solution is the same. Finally, a community data archive may already exist (e.g. EMPAIR for structural biology); in this case, the exact interface must be adapted to the existing technical solution, either by pushing the data through the archive interface or by attaching a dedicated Oneprovider with the archive.

\section{Implementation}
\label{sec:implementation}

Onedata4Sci solution consists of documentation, a \emph{fs2od} utility and the Onedata data management system itself. 

Utility \emph{fs2od} (filesystem to Onedata) is a CLI application written in Python 3. It monitors changes in a defined subtree of the filesystem, and according to the settings, it registers or updates the data within Onedata using the appropriate API calls.

\begin{figure*}[ht!]
\centering
\includegraphics[width=0.8\linewidth]{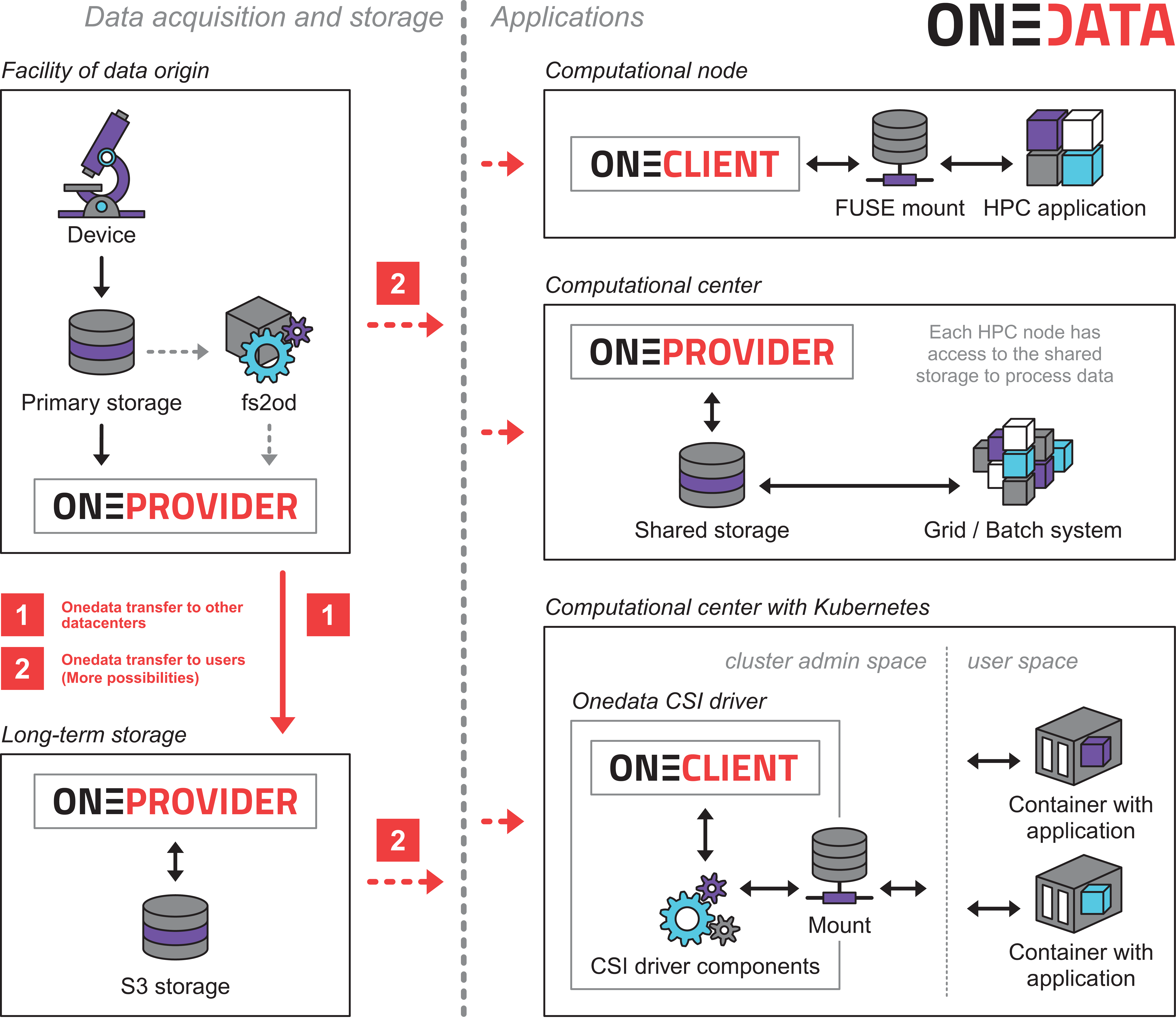}
\caption{The overview of Onedata4Sci ecosystem}
\label{fig:onedata4sci_full}
\end{figure*}

Following the architecture patterns described in the previous section, we now present the implementation details of Onedata4Sci (Figure~\ref{fig:onedata4sci_full}). Complete source codes are available on GitHub [\url{https://github.com/CERIT-SC/onedata4sci}]. Documentation and manual for installation are in Supplementary.

\subsection{Data acquisition}

Scientific experimental instruments are typically capable of storing the acquired data on a common (POSIX compatible) file system. The storage on which the data are created (stored from devices) is called primary storage. 

According to the parameters set in the application configuration file, the specified directories of the primary storage containing data sets are periodically searched at specified intervals. A subdirectory of a specified directory containing a metadata file of a defined name with arbitrary, even empty, content is considered a dataset. 

If a new dataset is discovered, the dataset registration workflow starts. This workflow consists of several steps to set up all the necessary requirements for storing and accessing the dataset. This includes, for example, the creation of an internal persistent dataset identifier (SpaceID), binding it to a specific location on the primary storage. 

In addition, replication policies and access permissions are set, and the access token is generated. A unique identifier is then written by the application to the metadata file for further use (e.g. for passing the data to the user by the operator or for loading into a laboratory information system). Depending on the settings, the application can also send the access information via a template-customisable email message to email addresses specified in the application configuration file (e.g. to instrument operators) or in the metadata file (to the addresses of users for whom a particular data set was captured). 

Loading the content of a dataset can be done either at once when it is created or continuously over a longer period of time. If a one-time loading is selected, the files and folders contained within the dataset are registered in the data management system when the dataset is created, and further changes are not reflected. This mode is suitable if the dataset is created all at once or if it is registered in the system only after the complete acquisition of the data. However, if the dataset is created over a longer period of time and it is also required to make the data available during the acquisition period, it is appropriate to allow continuous monitoring and updating of the dataset. Continuous content monitoring is enabled by the presence of a file with a defined name (e.g. empty ``.running'') in the root directory of the dataset. If this file is present, the corresponding Onedata functionality is enabled, and the monitoring and updating of the dataset content is performed. The continuous scanning parameters, e.g. the time delay between two consecutive scans, can be influenced in the \emph{fs2od} configuration file. When the signalling file is deleted, \emph{fs2od} disables continuous scanning. The reason for selectively enabling scanning is to reduce the number of unnecessary IO operations on the primary storage. In the case of a large number of ``live'' large datasets (terabytes of data in thousands of files), this scanning would already cause a considerable load on the storage. Continuous monitoring of changes can be de/activated for the entire Space. For this reason, we have chosen the approach of representing the dataset with just one Space. 

In some special cases, e.g. insufficient storage size for many datasets, the \emph{fs2od} utility allows the deletion of the whole dataset from the Oneprovider, called the primary provider, safely. This is only available when a replication mode is enabled. The utility ensures that all the data from a primary provider is transferred to at least one of the replicating providers. The dataset support at the primary provider is then ceased, and the space is deleted. This space is no longer synchronised, and all new data will not be uploaded. The user can also decide whether the data from a physical storage are deleted likewise. This is done by adjusting the configuration file.

In order to be consistent with FAIR principles, special care is also taken with any metadata contained in the metadata file. The metadata is loaded into the appropriate Onedata structures, thus allowing its further use, e.g. in search queries (Onedata allows searching using the attached Elasticsearch cluster) or its bulk export. 

When running, the \emph{fs2od} application must be able to access the dataset folders created on the primary storage in the same way as the Onedata Oneprovider component. If both \emph{fs2od} and Oneprovider run in containers, the paths to these folders within the containers must be identical. The two applications can run on the same compute node or, if the condition in the previous sentence is met, on different nodes (in which case the \emph{fs2od} application must have access to the Oneprovider's REST application interface). The data itself can be stored directly on the compute node or the disk array connected to it (Sect.~\ref{sssec:oneprovider_onsite}), or it can be mounted from elsewhere (e.g. by CIFS, Sect.~\ref{sssec:mounted_storage}). The chosen method and technology of data access will affect the achievable transfer rate.

The \emph{fs2od} application interacts with the central Onezone component and with one (primary) Oneprovider server or with other Oneprovider replication servers. The communication is realised through the REST application interface. The instance of the \emph{fs2od} application performs all operations under the identity of a specific (service) Onedata user. The authentication itself uses system-generated tokens sent along with the requests (in the HTTP header). 

\subsection{User access}

Once the user has an access token or a unique URL, they can access the dataset. The data is accessible to the user through the Onedata interface. Ordinary (non-IT) users can access the data most easily via web access. With this access, it is possible to work comfortably with datasets up to several GB. For larger data, we have developed a bulk download tool.

\subsubsection{Bulk download}
For bulk download, we provide a single-purpose application implemented in Python contained in a single file. This application allows the search of the entire directory structure. It supports downloading in multiple threads, resuming after an interrupted download. The number of used threads is customisable, either letting the end user not saturate his own connection or just allowing it.

\subsection{Computational processing}
\label{sec:implcomp}

Our target platform for computational processing is Kubernetes, an emerging \textit{de facto} standard for the execution of containerised applications. The data processing software is expected to be wrapped in a container image executable in the Kubernetes environment. For security reasons, running Oneclient (which requires FUSE kernel support, needing rather high container privileges consequently) in the application container is not acceptable in general. On the contrary, we leverage the native Kubernetes way to provide access to the data via a~well-defined \emph{Container Storage Interface} (CSI, \url{https://kubernetes.io/blog/2019/01/15/container-storage-interface-ga}). We developed a Onedata CSI driver. The driver is based on the existing \emph{csi-sshfs} project and is currently in the ``proof-of-concept'' phase (\url{https://github.com/CERIT-SC/csi-onedata}). The driver must be installed by the Kubernetes administrator, but the application containers can access Onedata spaces without the need for any higher privilege in turn.

\subsection{Archival}

After the datasets are captured and stored, it is desirable to deal with their permanent storage -- archiving. For this purpose, our solution offers the possibility of replicating the dataset to the storage sites designated for archiving. 

Replication of data to other repositories is implemented by setting the Onedata "Quality Of Service" functionality. For the dataset to be replicated, an instance of the additional Oneprovider service is added. Then, the actual data transfer is started. This workflow is described in Section~\ref{ssec:archival}. 

\subsection{Scalability and limitations}
\label{ssec:scalability}
During the development and testing of the proposed solution, we focused on scalability and possible limits. We did not encounter any limits arising from the size of the dataset or the number of files contained in it. A limiting factor may be a large number of users accessing the data simultaneously. This can be solved by increasing the number of compute bottlenecks running the Oneprovider service (see the Onedata documentation). 

Another factor affecting performance is the access speed of the chosen storage technology (data storage on NVMe drives vs. magnetic tapes) and how it is connected.

Furthermore, our solution is not primarily designed to work with datasets containing sensitive personal data. For example, it does not include functionality such as end-to-end encryption. However, if all the described system components run in a secure environment strictly separated from the Internet, the use of our solution is possible. 

\section{Results -- Use Cases}
\label{sec:results}

To show the applicability and versatility of our solution, we provide three distinct use cases used within the core facilities of Masaryk University. They differ in several aspects, from simple Onedata deployment to a complex custom-based solution. Table~\ref{tab:use_cases} presents an overview of these use cases.

\begin{table*}[ht!]
\centering
\begin{tabular}{llll}
\hline
                                     & \textbf{CF PLANTS} & \textbf{CF CELLIM} & \textbf{CF CryoEM} \\ \hline
\multicolumn{4}{l}{\textbf{Datasets overview}}                                                   \\ \hline
Number of datasets per year          & 20-30              &       500-1000& 300             \\
Average size of the dataset          & up to 1~GB         & about 500~GB& up to several TBs   \\
Average number of files per dataset  & up to 1000         & about 50& up to 20k      \\
Total amount of data stored          & hundreds of GBs    & 135 TB& a few PBs    \\ \hline
\multicolumn{4}{l}{\textbf{Architecture and user patterns}}                                      \\ \hline
Data acquisition                     & mounted storage    & mounted storage    & provider on site\\
User access                          & web interface      & web interface      & bulk download   \\
Dataset creators                     & CF operators       & CF users           & CF operators    \\ 
Data creation process                & all at once        & continual          & continual       \\ \hline
\end{tabular}
\caption{Overview of use cases}
\label{tab:use_cases}
\end{table*}

\subsection{Plant imaging data}
\label{ssec:plants}

Plant Sciences Core Facility (\url{https://plants.ceitec.cz}, CF PLANTS) provides access to cutting-edge infrastructure for plant and algae cultivation and environmental simulation and phenotyping analyses. One of the primary data sources is the phenotyping station (PlantSystem™, PSI, Drásov), which provides information, such as the plant’s morphology architecture, fluorescence and colour characterisation. In this case, the data is just an automatically generated time-based spreadsheet with measured properties and images of the cultivated plants in time. Therefore, the space requirements for such a dataset are minimal, reaching up to several hundred megabytes at most.

Since the device is operated solely by the core facility employees (operators), the creation of the datasets is limited to only a few people. Due to this, the Onedata solution deployment is relatively straightforward. On the dedicated network storage, a directory is allocated and connected using CIFS to the personal computers of the individual CF operators as a shared drive. Within the drive, each newly created directory with a corresponding YAML metadata file is registered by the \emph{fs2od} daemon running on the network storage. After that, the operator emails the public web link to a particular user. Read-only web access is sufficient in this case, as the data are relatively small, and further processing might be done locally. It is worth noting that in the future, automation of this final step using the user’s email address from the metadata YAML could be facilitated by \emph{fs2od}.

\subsection{Cellular Imaging data}
\label{ssec:cellim}
Core Facility Cellular Imaging (\url{https://cellim.ceitec.cz}, CF CELLIM) provides light microscopy services and comprehensive expertise in sample preparation, image acquisition and data analysis. At the time of writing, CELLIM operates 11 microscopes in total. All microscopes are connected to an intranet through dedicated Windows computers with specialised acquisition software and are also connected to the university storage system with more than 5 PB space for the data.  Furthermore, CF operates a local Windows server machine with 260 terabytes of disk space with license-restricted software that can be used for post-processing and analysis of the acquired data. The CF operators assign access to the server manually and assign a size quota, in general, 1~TB per user. Based on the size of the project, this quota can be adjusted to cover the needs of users, especially for light-sheet microscope users. The individual datasets might span several orders of magnitude, ranging from a few megabytes to several terabytes, so the local server is unsuitable for long-term storage. For long-term storage data, users have to find capacity, mainly on university storage server. 

The Oneprovider service runs on a separate virtual node with a Linux operating system because the requirements for running the Oneprovider service prescribe Linux as the guest operating system. A root directory with datasets from the lab server is mounted on this virtual node using the CIFS protocol.

\subsection{Cryo-electron microscopy data}
\label{ssec:cryoem}
The cryo-electron microscopy core facilities provide access to electron microscopy instrumentation for structural and cellular biology research. This data is the most complex of the three use cases, both in terms of the amount of data stored (over 1~PB) and the number of files. However, \emph{Onedata4Sci} is suitable even here. We successfully tested the solution in one shared laboratory at Masaryk University.

The laboratory has developed its own LIMS (Laboratory Information Management Systems) that can be used to manage the entire process of sample preparation and scanning. 

The laboratory operates a high-capacity disk array. The Oneprovider component and the \emph{fs2od} application run directly on this disk array. In this way, direct access to the data is ensured. The LIMS operator of the lab fills in the metadata of the dataset, after which the metadata file is stored in the directory where the acquired (raw) data are stored. The microscope data itself is then uploaded to the prepared directory with the saved metadata file.

The Oneprovider service itself runs in a Docker container. The necessary directories are mounted to the container. This ensures that the Oneprovider environment is isolated from the disk array front node environment. 

\section{Conclusions and future work}
\label{sec:conclusions}

In this article, we present \emph{Onedata4Sci}, a data management solution based on the Onedata data management system. \emph{Onedata4Sci} is focused on the management of life science data, and it covers four key steps in the management of this data, i.e. data acquisition, user access, computational processing and archiving. We focused on finding a reasonable tradeoff between tightly tailored single-purpose solutions on one side and very generic (but difficult-to-use out-of-box) ones. The applicability of \emph{Onedata4Sci} is shown in three distinct use cases -- plant imaging data, cellular Imaging data, and cryo-electron microscopy data. Despite the use cases covering very different types of data and user approaches, they exhibit common patterns we could focus on. \emph{Onedata4Sci} demonstrates an ability to successfully handle all these requirements. All the use cases are deployed in production, handling the true experimental data, and the users are starting to rely on the service in their routine work. Complete source codes of \emph{Onedata4Sci} are available on GitHub (\url{https://github.com/CERIT-SC/onedata4sci}), and its documentation and manual for installation are also provided.

The management of scientific data is indeed a long-term task. Therefore, we intend to support and further improve \emph{Onedata4Sci} as the individual CFs depend on it. In future, we plan to design a web portal that will facilitate the management of individual data sets, add an overview of user datasets and allow the search based on the metadata and other various properties. 

\section{Acknowledgement}
\label{sec:acknowledgement}
We acknowledge the core facility CELLIM supported by the Czech-BioImaging large RI project (LM2023050 funded by MEYS CR) and Biological Data Management and Analysis Core Facility of CEITEC Masaryk University, funded by ELIXIR CZ research infrastructure (MEYS Grant No: LM2023055) for their support with obtaining scientific data presented in this paper. Plant Sciences Core Facility of CEITEC Masaryk University is acknowledged for the technical support.
\section{Funding}
\label{sec:funding}

This research received backing from the "Archival and Sharing of Scientific Data in Onedata" grant (FR CESNET: 696/2022) awarded to Tomáš Svoboda.

 \bibliographystyle{elsarticle-num} 
 \bibliography{references}

\end{document}